\documentstyle[aps,prl,twocolumn, epsf]{revtex}

\def\be{\begin{equation}}
\def\ee{\end{equation}}
\def\bea{\begin{eqnarray}}
\def\eea{\end{eqnarray}}
\def\bma{\begin{mathletters}}
\def\ema{\end{mathletters}}
\def\H{{\cal H}}

\def\D{{\cal D}}
\def\N{{\cal N}}
\def\T{{\cal M}}

\def\ide{I}
\def\tr{{\rm tr}}
\def\rhot{\rho\pt}
\def\pt{^{T_A}}
\def\norm#1{\Vert#1\Vert}
\def\normt#1{\norm{#1}_1}
\def\idty{{\rm 1\mkern -5.4mu I}}

\def\C{\hbox{\kern.3em\vrule
     height 1.5ex depth -.1ex width .2pt\kern-.3em\rm C}}
\newcommand{\bra}[1]{\mbox{$\langle #1 |$}}
\newcommand{\ket}[1]{\mbox{$| #1 \rangle$}}

\newcommand{\proj}[1]{\ket{#1}\!\bra{#1}}
\def\ketbra#1#2{\ket{#1}\!\bra{#2}}
\def\abs#1{\vert#1\vert}
\def\flip{{\bf F}}

\begin{document}
\title{A computable measure of entanglement}
\author{G. Vidal \\
Institut f\"{u}r Theoretische Physik,
 Universit\"{a}t Innsbruck,\\
 Technikerstra\ss e 25
 A-6020 Innsbruck, Austria \\
and R.F. Werner\\
 Institut f\"{u}r Mathematische Physik,
 TU Braunschweig,\\
 Mendelssohnstr. 3,
 38304 Braunschweig, Germany}
\maketitle

\begin{abstract}
We present a measure of entanglement that can be computed
effectively for any mixed state of an arbitrary bipartite system.
We show that it does not increase under local manipulations of the
system, and use it to obtain a bound on the teleportation capacity
and on the distillable entanglement of mixed states.

\end{abstract}

\section{Introduction}

In recent years it has been realized that quantum mechanics offers
unexpected possibilities in information transmission and
processing, and that quantum entanglement of composite systems
plays a major role in many of them. Since then, a remarkable
theoretical effort has been devoted both to classifying and
quantifying entanglement.

Pure-state entanglement of a bipartite system is presently
well-understood, in that the relevant parameters for its optimal
manipulation under local operations and classical communication
(LOCC) have been identified, in some asymptotic sense \cite{Be96}
as well as for the single copy case \cite{Ni99}. Given an
arbitrary bipartite pure state $\ket{\psi_{AB}}$, the {\em entropy
of entanglement} $E(\psi_{AB})$ \cite{Be96}, namely the
von-Neumann entropy of the reduced density matrix $\rho_A \equiv
\mbox{Tr}_B \proj{\psi_{AB}}$, tells us exhaustively about the
possibilities of transforming, using LOCC, $\ket{\psi_{AB}}$ into
other pure states, in an asymptotic sense. When manipulating a
single copy of $\ket{\psi_{AB}}$, this information is provided by
the $n$ entanglement monotones $E_{l}=\sum_{i=l}^n \lambda_i~$
($l=1,...,n$) \cite{Ni99}, where $\lambda_i$ are the eigenvalues
of $\rho_A$ in decreasing order.

Many efforts have also been devoted to the study of mixed-state
entanglement. In this case several measures have been proposed.
The entanglement of formation $E_F(\rho)$ \cite{Be962} ---or, 
more precisely, its renormalized version, the entanglement cost $E_C(\rho)$ \cite{Ha00}--- and the distillable
entanglement $E_D(\rho)$ \cite{Be962} quantify, respectively, the
asymptotic pure-state entanglement required to create $\rho$, and
that which can be extracted from $\rho$, by means of LOCC. The
relative entropy of entanglement \cite{Ve98} appears as a third,
related measure \cite{Ve00} which interpolates between $E_C$ and
$E_D$ \cite{Ho00}.

However, in practice it is not known how to effectively compute
these measures, nor any other, for a generic mixed state, because
they involve variational expressions. To our knowledge, the only
exceptions are Wootter's closed expression for the entanglement of
formation $E_F(\rho)$ [and {\em concurrence} $C(\rho)$] of
two-qubit states \cite{Wo98}, and its single-copy analog
$E_2(\rho)$ also for two qubits \cite{Vi00}.

Multipartite pure-state entanglement represents the next order of
complexity in the study of entanglement, and is of interest,
because one hopes gain a better understanding of the correlations
between different registers of a quantum computer. Consider a
tripartite state $\ket{\psi_{ABC}}$. Some of its entanglement
properties depend on those of the two-party reduced density
matrices, which are in a mixed state. For instance,  the relative
entropy of $\rho_{AB}\equiv$ tr$_C\proj{\psi_{ABC}}$ has been
used to prove that bipartite and tripartite pure-state
entanglement are asymptotically inequivalent \cite{Li99}. Thus,
the lack of an entanglement measure that can be easily computed
for bipartite mixed states is not only a serious drawback in the
study of mixed-state entanglement, but also a limitation for
understanding multipartite pure-state entanglement.

The aim of this paper is to introduce a {\em computable} measure
of entanglement \cite{foot}, and thereby fill an important gap in the study of entanglement.
 It is based on the trace norm of the partial transpose $\rhot$ of
the bipartite mixed state $\rho$, a quantity whose evaluation is
completely straightforward using standard linear algebra packages.
It essentially measures the degree to which $\rhot$ fails to be
positive, and therefore it can be regarded as a quantitative
version of Peres' criterion for separability \cite{Peres}. From
the trace norm of $\rhot$, denoted by $||\rhot||_1$, we will
actually construct two useful quantities. The first one is the
{\em negativity},
 \be
   \N(\rho) \equiv  \frac{ \normt{\rho\pt}-1}{2},
 \ee
which corresponds to the absolute value of the sum of negative eigenvalues of
$\rhot$ \cite{others}, and which vanishes for unentangled states. As
we will prove here, $\N(\rho)$ does not increase under LOCC, i.e.,
it is an entanglement monotone \cite{monotone}, and as such it can
be used to quantify the degree of entanglement in composite
systems. We will also consider the {\em logarithmic negativity},
 \be
   E_{\N}(\rho) \equiv \log_2||\rhot||_1, \ee
 which again exhibits some
form of monotonicity under LOCC (it does not increase during deterministic distillation protocols) and is, remarkably, an additive quantity.

 The importance of $\N$ and $E_{\N}$ is boosted, however, beyond their practical computability by two results that link theses measures with relevant parameters characterizing entangled mixed states.
 The negativity will be shown to
bound the extent to which a single copy of the state $\rho$ can be
used, together with LOCC, to perform quantum teleportation
\cite{tele}. In turn, the logarithmic negativity bounds the
distillable entanglement $E_D^{\epsilon}$ contained in $\rho$, that
is, the amount of ``almost pure''-state entanglement that can be
asymptotically distilled from $\rho^{\otimes N}$, where ``almost''
means that some small degree $\epsilon$ of imperfection is allowed
in the output of the distillation process. 

 Remarkably, this last result has already found an application in the context of asymptotic transformations of bipartite entanglement \cite{Vidal}, as a means to prove that (PPT) bound entangled states \cite{Ho98} cannot be distilled into entangled pure states even if loaned (i.e. subsequently recovered for replacement) pure-state entanglement is used to assist the distillation process. In this way, the bound on distillability implied by $E_{\N}$ has contributed to prove that, in a bipartite setting, asymptotic local manipulation of mixed-state entanglement is sometimes, in contrast to its pure-state counterpart, an inherently irreversible process.

 We have divided this paper into VII sections. In section II some properties of the negativity $\N$, such as its monotonicity under LOCC, and of the logarithmic negativity $E_{\N}$ are proved. We also discuss a more general construction leading to several other (non-increasing under LOCC) negativities. In sections III and IV we derive, respectively, the bounds on teleportation capacity and on asymptotic distillability. Then in section V we calculate the explicit expression of $\N$ and $E_{\N}$ for pure states and for some highly symmetric mixed states, also for Gaussian states of light field. In section VI extensions of these quantities to multipartite systems are briefly considered, and section VII contains some discussion and conclusions.

\section{Monotonicity of $\N(\rho)$ under LOCC.}

In this section we show that the negativity $\N(\rho)$ is an
entanglement monotone. We first give a rather detailed proof of
this result. Then we sketch an argument extending this observation
to several other similarly constructed negativities---like the
{\em robustness of entanglement} \cite{rob}.

\subsection{Definition and basic properties}

From now on we will denote by $\rho$ a generic state of a
bipartite system with finite dimensional Hilbert space
$\H_A\otimes \H_B \equiv\C^{d_A}\otimes \C^{d_B}$ shared by two
parties, Alice and Bob. $\rhot$ denotes the {\em partial
transpose} of $\rho$ with respect to Alice's subsystem, that is
the hermitian, trace-normalized operator defined to have matrix
elements,
 \be
    \bra{i_A , j_B}\rhot \ket{ k_A , l_B}
  \equiv \bra{k_A ,j_B} \rho \ket{i_A ,l_B},
\label{transpose}
 \ee
 for a fixed but otherwise arbitrary
orthonormal product basis $\ket{i_A ,j_B} \equiv \ket{i}_{A}
\otimes \ket{j}_{B} \in \H_A\otimes \H_B$. The {\em trace norm}  of
any hermitian operator $A$ is $\normt{A} \equiv \tr \sqrt{A^{\dagger}A}$ \cite[Sec.VI.6]{ReedSimon}, which is equal to the sum of the
absolute values of the eigenvalues of $A$, when $A$ is hermitian \cite{dagger}.
For density matrices all eigenvalues are positive and thus
$\normt{\rho} = \tr\rho = 1$. The partial transpose $\rhot$ also
satisfies $\tr[\rhot]=1$, but since it may have negative
eigenvalues $\mu_i < 0$, its trace norm reads in general
 \be
  \normt{\rhot} = 1 + 2|\sum_i \mu_i| \equiv 1 + 2\N(\rho).
 \ee
Therefore the negativity $\N(\rho)$ ---the sum $|\sum_i\mu_i|$ of the negative
eigenvalues $\mu_i$ of $\rhot$---measures by how much $\rhot$
fails to be positive definite. Notice that for any separable or
unentangled state $\rho_s$ \cite{werner},
 \be
\rho_s = \sum_k p_k \proj{e_k, f_k}; ~~~~~ p_k \geq 0, \sum_k p_k = 1,
\label{sepa}
\ee
its partial transposition is also a separable state  \cite{Peres},
\be
\rho_s^{T_A} = \sum_k p_k \proj{e_k^*, f_k} \geq 0,
\label{sepatra}
\ee
and therefore $\normt{\rhot_s}=1$ and $\N(\rho_s)=0$.

The practical computation of $\N(\rho)$ is straightforward, using
standard linear algebra packages for eigenvalue computation of
hermitian matrices. On the other hand, this representation is not
necessarily the best for proving estimates and general properties
of $\N(\rho)$. To begin with a simple example, consider the
property that $\N(\rho)$ does not increase under mixing:

{\bf Proposition 1: \it $\N$ is a convex function, i.e.,
 \be \N\left(\sum_ip_i\rho_i\right)
        \leq\sum_ip_i\N\left(\rho_i\right)\;,
 \ee
whenever the $\rho_i$ are hermitian, and $p_i\geq0$ with
$\sum_ip_i=1$.}

There is nothing to prove here, when we write
$\N(\rho)=(\normt{\rho\pt}-1)/2$, and observe that $\normt\cdot$,
like any norm, satisfies the triangle inequality and is
homogeneous of degree $1$ for positive factors, hence convex.

However, the fact that $\normt\rho$ is indeed a norm is not so
obvious, when it is defined in terms of the eigenvalues. This is
shown best by rewriting it as a variational expression. Our reason
for recalling this standard observation from the theory of the
trace norm is that the same variational expression will be crucial
for showing monotonicity under LOCC operations. The variational
expression is simply the representation of a general hermitian
matrix $A$ as a {\it difference} of positive operators: Since we
are in finite dimension we can always write
\be
   A =  a_+\rho^+ - a_-\rho^-,
  \label{decomp}
 \ee
where $\rho^{\pm}\geq 0$ are density matrices ($\tr[\rho^{\pm}] =
1$) and $a_{\pm} \geq 0$ are positive numbers. Note that by taking
the trace of this equation we simply have $\tr[A]=a_+-a_-$.

{\bf Lemma 2}: For any hermitian matrix $A$ there is a
decomposition of the form (\ref{decomp}) for which $a_++a_-$ is
minimal. For this decomposition, $\normt A=a_++a_-$, and $a_-$ is
the absolute sum of the negative eigenvalues of $A$.

{\em Proof}:
 Let $P^-$ be the projector onto the negative
eigenvalued subspace of $A$, and $\N = -\tr[AP^-]$ the absolute sum
of the negative eigenvalues. We can reverse the decomposition
(\ref{decomp}) to obtain that $A+a_-\rho^-$ is positive
semidefinite. This implies that
 \be
   0 \leq \tr[(A+a_-\rho^-)P^-] = -\N + a_-\tr[\rho^-P^-]
 \ee
But $\tr[\rho^-P^-]\leq 1$, that is $a_- \geq \N$. This bound can
be saturated with the choice $a_-\rho^- \equiv -P^- AP^-$
(corresponding to the Jordan decomposition of $A$, where $\rho^-$
and $\rho^+$ have disjoint support), which ends the proof. $\Box$

For the negativity we therefore get the formula
 \be
   \N(A) = \inf\Bigl\{a_-  \Big\vert
                  A\pt=a_+\rho^+-a_-\rho^-\Bigr\}\;,
 \label{neg1}
 \ee
where the infimum is over all density matrices $\rho^\pm$ and
$a_\pm\geq0$.

Another remarkable property of $\N(\rho)$ is the easy way in which
$\N(\rho_1\otimes\rho_2)$ relates to the negativity of $\rho_1$
and that of $\rho_2$. This relationship is an important, but
notoriously difficult issue for discussing asymptotic properties
of entanglement measures (see, e.g., \cite{VW} for a discussion
and a counterexample to the conjectured additivity of the relative
entropy of entanglement).

For the entanglement measure proposed in this paper we get
additivity for free. We start from the identity
$\normt{\rho_1\otimes\rho_2}=\normt{\rho_1}\;\normt{\rho_2}$,
which is best shown by using the definition of the trace norm via
eigenvalues, and  observe that partial transposition commutes with
taking tensor products. After taking logarithms, we find for the
logarithmic negativity:
 \be\label{addi}
   E_\N(\rho_1\otimes\rho_2)=E_\N(\rho_1)+E_\N(\rho_2)\;.
 \ee
 It might seem from this that $E_\N$ is a candidate for the much
sought for canonical measure of entanglement. However, it has
other drawbacks. For instance, it is {\it not} convex, as is
already suggested by the combination of a convex functional (the
trace norm) with the concave log function, which implies that it increases under some LOCC. And although it has an interesting, monotonic behavior during asymptotic distillation (as shown in section IV), it does not correspond to the entropy of entanglement for pure states (see section V).

\subsection{Negativity as a mixed-state entanglement monotone.}

By definition, a LOCC operation (possibly for many parties)
consists of a sequence of steps, in each of which one of the
parties performs a local measurement, and broadcasts the result to
all other parties. In each round the local measurement chosen is
allowed to depend on the results of all prior measurements. If at
the end of a LOCC operation with initial state $\rho$ the
classical information available is ``$i$'', which occurs with
probability $p_i$, and final state conditional on this occurrence
is $\rho'_i$, we require of an entanglement monotone \cite{monotone} $E$ that
\begin{equation}\label{Emono}
  E(\rho)\geq\sum_ip_iE(\rho'_i)\;.
\end{equation}
It is clear by iteration that this may be proved by looking at
just one round of a LOCC protocol, consisting of a single local
operation. In the present case, since $\N$ makes no distinction
between Alice and Bob, it suffices to consider just one local
measurement by Bob.

Now the most general local measurement is described by a family
$\T_i$ of completely positive linear maps such that, in the
notation used in the previous paragraph, $\T_i(\rho)=p_i\rho'_i$.
These maps satisfy the normalization condition
$\sum_i\tr(\T_i(\rho))=\tr(\rho)$. This can be further simplified
\cite{monotone} when some $\T_i$ can be decomposed further into
completely positive maps, e.g., $\T_i=\T'_i+\T''_i$. Then we may
simply consider the finer decomposition as a finer measurement,
with the result $i$ replaced by two others,  $i'$ and $i''$. Using
the convexity already established it is clear that it suffices to
prove (\ref{Emono}) for the finer measurement. That is, we can
assume that there are no proper decompositions of the $\T_i$, or
that $\T_i$ is ``pure''. This is equivalent to $\T_i$ taking pure
states to pure states, or to the property \cite{invite} that it
can be written with a single Kraus summand. Taking into account
that this describes a local measurement by Bob, we can write
\begin{equation}\label{Ti}
  \T_i(\rho)=(\ide_A\otimes M_i)~\rho~
  (\ide_A\otimes M_i^{\dagger})\;,
\end{equation}
 where the Kraus operators $M_i$ must satisfy the normalization
condition $\sum_{i} M^{\dagger}_iM_i \leq I_B$. For computing the
right hand side of Eq.~(\ref{Emono}) we need that
\begin{equation}\label{miTA}
  \T_i(\rho)\pt=\T_i\Bigl(\rho\pt\Bigr)\;,
\end{equation}
 which immediately follows from (\ref{Ti}) by expanding $\rho$ as
a sum of (not necessarily positive) tensor products. A similar
formula holds for Alice's local operations, but with a modified
operation $\T_i$ on the right hand side, in which the Kraus
operators have been replaced by their complex conjugates.

Consider the decomposition
 \be
\rho\pt = (1\!+\!N)\rho^+ - N\rho^-. \label{decomprho}
 \ee
with density operators $\rho^\pm$ and $N=\N(\rho)$. Then we can
also decompose the partially transposed output states:
 \bea
  p_i(\rho'_i)\pt
    &=& \T_i(\rho)\pt=\T_i(\rho\pt)\nonumber\\
    &=&(1\!+\!N)\T_i(\rho^+)- N\T_i(\rho^-).
 \label{decomprhoi}
 \eea
Dividing by $p_i$ we get a decomposition of precisely the sort
(\ref{neg1}) defining $\N(\rho'_i)$. The coefficient $a_-=N/p_i$
must be larger than the infimum, i.e., $\N(\rho_i')\leq N/p_i$.
Multiplying by $p_i$ and summing, we find the following inequality.

{\bf Proposition 3:}
 \be
    \sum_i p_i \N(\rho_i')\leq\N(\rho) ,
 \ee
i.e., $\N(\rho)$ is indeed an entanglement monotone.

\subsection{Other negativities.}

Both the proof of convexity and the proof of monotonicity are
based on the variational representation of the trace norm in Lemma
2. The abstract version of this Lemma is the definition of the
so-called {\it base norm} $\norm\cdot_S$ associated with a compact
set $S$ in a real vector space \cite{Nagel}. The negativity
introduced above then corresponds to a special choice of $S$, and
we can easily find the property of $S$ required for proving LOCC
monotonicity in the abstract setting. Other choices of $S$ then
lead to other entanglement monotones, some of which have been
proposed in the literature.

 For our purposes we can take $S$ as an arbitrary compact convex subset
of the hermitian operators with unit trace, whose real linear hull
equals all hermitian operators. Then, in analogy to Lemma 2, we
define the associated base norm and ``$S$-negativity'' as
 \bea\label{normS}
   \norm{A}_S&=&\inf\Bigl\{a_++a_-  \Big\vert
                  A=a_+\rho^+-a_-\rho^-,\Bigr.   \\
   &&\hskip80pt\Bigl.
                a_\pm\geq0,\ \rho^\pm\in S\Bigr\},\nonumber\;\\
   \N_S(A)&=&\inf\Bigl\{a_-  \Big\vert \label{NS}
                  A=a_+\rho^+-a_-\rho^-,\Bigr.   \\
   &&\hskip80pt\Bigl.
                a_\pm\geq0,\ \rho^\pm\in S\Bigr\},\nonumber\;\\
 \eea
Note that once again, if $A$ has trace $1$ we have that $\norm{A}_S =
1+2\N_S(A)$. Then norm and convexity properties of $\N_S$ and
$\norm\cdot_S$ follow exactly as before.

Taking $S$ as the set of all density matrices, we get $\norm
A_S=\normt A$, for all hermitian $A$, and a totally uninteresting
entanglement quantity, as $\N_S(\rho)$ vanishes for all density
matrices.  The negativity of the previous section corresponds to
the choice of $S$ equal to the set of all matrices $A$ such that
$A=A^\dagger$, $\tr A=1$, and $A\pt\geq0$ (additionally, we have replaced $A^{T_A}$ with $A$ in the LHS of eq. (\ref{neg1}) $A$, so that we can write $\N(\rho)$ instead of $\N(\rho^{T_A})$).

We could have also taken $S$ as the subset of density matrices
with positive partial transpose, $\rho^{\pm}\geq 0$ and $\rho^{\pm
T_A}\geq 0$. In this case $S$ corresponds to all states such that
its partial transpose is also a state. The resulting quantity we
will denote by $\N_{ppts}$. Even more restrictively, if we take
for $S$ the set of {\it separable} density operators, i.e. we take
$\rho^{\pm}$ (and therefore also $\rho^{\pm T_A}$) in eqs.
(\ref{normS},\ref{NS}) to be separable, the corresponding quantity
$\N_{ss}$ amounts to the {\em robustness of entanglement},
originally introduced in \cite{rob} (see also \cite{Rudolph}) as
the minimal amount of separable noise needed to destroy the
entanglement of $\rho$. From the inclusions between the respective
sets $S$ we immediately get the inequalities
 \be\label{NNN}
   \N_{ss}(\rho)\geq\N_{ppts}(\rho)\geq\N(\rho)\geq0\;.
 \ee
In general, all these inequalities are strict. For example,
$\N_{ss}(\rho)$ vanishes only on separable states, whereas
$\N_{ppts}(\rho)$ and $\N(\rho)$ vanish for all ppt-states.

We claim that also $\N_{ss}$ and $\N_{ppts}$ are entanglement
monotones. The proof is quite simple. An analysis of the arguments
given in the previous section shows that we really used only one
property of $S$, namely that for all operations $\T_i$ appearing
in a LOCC protocol, we have $\T_i(\rho)\in S^\vee$, whenever
$\rho\in S^\vee$, where $S^\vee$ notes the cone generated by $S$
(equivalently the set of $\lambda \rho$ with $\lambda\geq0,\rho\in
S$). But this is obvious for both separable states and ppt-states.

\section{Upper bound to teleportation capacity}

 Sections III and IV are devoted to discuss applications of the previous results. More specifically, we derive bounds to some properties characterizing the entanglement both of a single copy of a mixed state $\rho$ (this section) and of asymptotically many copies of it (next section).

For a single copy of a bipartite state $\rho$ acting on
$\C^{d_1}\otimes\C^{d_2}$, where we set $d_1= d_2 \equiv m$ for
simplicity, an important question in quantum information theory
is to what extend this state can be used to implement some given
tasks requiring entanglement, such as teleportation. The best
approximation $P_{opt}(\rho)$ to a maximally entangled state,
 \be
  \ket{\Phi^+} \equiv \frac{1}{\sqrt{m}} \sum_{\alpha=1}^m \ket{\alpha_A\otimes \alpha_B},
 \ee
that can be obtained from $\rho$ by means of LOCC is then
interesting, because it determines, for instance, how useful the
state $\rho$ is to {\em approximately} teleport $\log_2 m$ qubits
of information. In this section we will show that the negativity
$\N(\rho)$ provides us with an explicit lower bound on how close
$\rho$ can be taken, by means of LOCC, to the state
$\Phi^+$. From here a lower bound on the teleportation
distance (i.e., an upper bound on how good teleportation results from
$\rho$) will also follow.

\subsection{Singlet distance}

In order to characterize the optimal state $P_{opt}(\rho)$
achievable from $\rho$ by means of LOCC, we need to quantify its
closeness to the maximally entangled state
$P_+\equiv\proj{\Phi^+}$. Let $\rho_1$ and $\rho_2$ be two density
matrices. The trace norm of $\rho_1\!-\!\rho_2$, (or {\em absolute
distance} \cite{nielsen}), is a measure of the degree of
distinguishability of $\rho_1$ and $\rho_2$, and it is therefore
reasonable to use it to measure how much $P(\rho)$---the state
resulting from applying a local protocol $P$ to state
$\rho$---resembles $P_+$. In what follows we will prove that the
negativity is a lower bound to the {\em singlet distance} of
$\rho$,
 \be
\Delta (P_+,\rho) \equiv \inf_P ||P_+ - P(\rho) ||_1,
\label{singletdist}
\ee
where the infimum is taken over local protocols $P$.

We start be recalling that the absolute distance $D(\rho_1,\rho_2)
\equiv||\rho_1-\rho_2||_1$ is a convex function \cite{nielsen},
 \be
\sum_i p_i D(\sigma, \rho_i) \geq D(\sigma,\sum_i p_i \rho_i),
\label{prop1} \ee
 which confirms, as already assumed, that the
optimal approximation $P(\rho)$ to $P_+$ can always be chosen to
be a single state---as opposed to a distribution of states $\{p_i,
\rho_i\}$ corresponding to the output of a probabilistic
transformation. Therefore, in eq. (\ref{singletdist}) we need only
consider {\em deterministic} protocols $P$ based on LOCC.

 A second feature of the absolute distance that we need is that
\be
D(W\rho_1W^{\dagger},W\rho_2W^{\dagger})= D(\rho_1,\rho_2),
\label{prop2} \ee
 for any unitary transformation $W$. Properties
(\ref{prop1}) and (\ref{prop2}) together imply that the best
approximation to the maximally entangled state $P_+$ can always be
``twirled'' without losing optimality. Consider the state
 \be
\int dU\  U\otimes U^* P_{opt}(\rho) U^{\dagger}\otimes U^{\dagger
*}, \label{twirling} \ee
 which the parties can locally obtain from
$P_{opt}(\rho)$ by Alice applying an arbitrary unitary $U$, by Bob
applying $U^*$, and by then deleting the classical information
concerning which unitary has been applied. It follows from the
invariance of $P_+$ under $U\otimes U^*$ and from property
(\ref{prop2}) that $D(U\otimes U^* P_{opt}(\rho)U^{\dagger}\otimes
U^{\dagger *},P_+)= D(P_{opt}(\rho), P_+)$ for any $U$. Then
property (\ref{prop1}) implies that the mixture in eq.
(\ref{twirling}) is not further away from $P_+$ than
$P_{opt}(\rho)$. But $P_{opt}(\rho)$ was already minimizing eq.
(\ref{singletdist}), and therefore state (\ref{twirling}) must
also be optimal.

We can then assume that $P_{opt}(\rho)$ has already undergone a
twirling operation. This means that it is a {\em noisy singlet}
\cite{Horox3-99},
 \be\label{noising}
     \rho_p = pP_+\;+\; (1-p) \frac{I\otimes I}{m^2},
 \ee
 from which the
absolute distance to $P_+$ can be easily computed, $D(P_+,\rho_p)
= 2(1-p)(m^2-1)/m^2.$ Similarly, the trace norm of $\rho_p^{T_A}$
reads $||\rho_p^{T_A}||_1 = mp+(1-p)/m$, and therefore
\be
D(P_+,\rho_p) = 2 (1-\frac{||\rho_p^{T_A}||_1}{m}).
 \ee
 The lower
bound to the singlet distance (\ref{singletdist}) follows now
straightforwardly from the monotonicity of $||\rho^{T_A}||_1$ (or
$\N(\rho)$) under LOCC, that is, $||\rho^{T_A}||_1 \geq
||P_{opt}(\rho)^{T_A}||_1$, and therefore
\be
\Delta(P_+,\rho) \geq 2(1-\frac{||\rho^{T_A}||_1}{m}).
\label{bound-singlet}
\ee
Therefore we have proved the following bound for the singlet distance.

{\bf Proposition 4}:
\be
\Delta(P_+,\rho) \geq 2(1-\frac{1+2\N(\rho)}{m}).
\ee

\subsection{Teleportation distance}

A quantum state $\rho$ shared by Alice and Bob can be used as a
teleportation channel $\Lambda$ \cite{tele}. That is, given the
shared state $\rho$ and a classical channel between the parties,
Alice can transmit an arbitrary (unknown) state $\phi \in {\cal
C}^m$ to Bob with some degree of approximation. Let
$\Lambda_{T,\rho}(\phi)$ be the state that Bob obtains when Alice
sends $\phi$ using $\rho$ and some protocol $T$ involving LOCC
only. The teleportation distance
 \be
    d(\Lambda) \equiv \int d\phi\ D(\phi,\Lambda(\phi)),
 \ee
where $D(\phi,\Lambda(\phi))\equiv ||\proj{\phi} -
\Lambda(\phi)||_1$, can be used to quantify the degree of
performance of the channel. The measure $d\phi$ is consistent with the
Haar measure $dU$ in $SU(m)$, and thus $d(\Lambda)$ is invariant
under the twirling of the channel, that is the application of an
arbitrary unitary $U$ to $\phi$ previous to the teleportation,
followed by the application of $U^\dagger$ after the teleportation
scheme. Indeed,
 \be
   d(\Lambda) = \int dW\ D(WP_0W^{\dagger},\Lambda(WP_0W^{\dagger})),
  \label{distance2}
 \ee
for some reference state $P_0\equiv\proj{\phi_0}$, and using
property (\ref{prop2}) of the trace norm, eq. (\ref{distance2}) is
also equal to
 \be
   \int dW\
d(WP_0W^{\dagger},U^{\dagger}\Lambda(UWP_0W^{\dagger}U^{\dagger})U).
 \ee
We can now average over $U$ to obtain
 \be
   d(\Lambda) = \int dU\ \int d\phi D(\phi, U^{\dagger}\Lambda(U\phi)U),
 \ee
 where the right side of the equation corresponds to the
teleportation distance of the twirled channel.

We next adapt a reasoning of the Horodeckis \cite{Horox3-99} to
our present situation. It uses an isomorphism between states
$\rho_{\Lambda}$ and channels $\Lambda$ due to Jamio{\l}kowski
\cite{Jami}. Let us ascribe the channel $\Lambda$ to the state
$\rho_{\Lambda} = (I\otimes \Lambda) P_+$. The state
$\rho_{\Lambda}$ can be produced by sending Bob's part of the
bipartite system in state $P_+$ down the channel $\Lambda$.
Conversely, the standard teleportation protocol \cite{tele} (or a
slight and obvious modification of it) applied to state
$\rho_{\Lambda}$ reproduces the channel $\Lambda$ with probability
$1/m^2$. However, if the state $\rho_{\Lambda}$ is a noisy singlet
$\rho_p$, then the corresponding channel is the depolarizing
channel
 \be
\Lambda_p^{dep}(\varrho) = p\varrho + (1-p)\frac{I}{m},
 \ee
 which the standard teleportation scheme reproduces with {\em certainty}
using state $\rho_p$. For this case $d(\Lambda_p^{dep}) =
2(1-p)(m-1)/m$. Therefore there is a complete physical equivalence
between noisy singlets and depolarizing teleportation channels. In
addition,
 \be
d(\Lambda_p^{dep}) = \frac{m}{m+1} D(P_+,\rho_p).
\label{relation}
\ee
 Now, since both quantities $d$ and $D$ are invariant under
twirling, and any channel (state) can be taken into the
depolarizing (noisy singlet) form, this equality holds for any
channel $\Lambda$ and state $\rho_{\Lambda}$.

{\bf Lemma 5} (adapted from \cite{Horox3-99}): The minimal
distance $d_{min}(\rho)$ that can be achieved when using the
bipartite state $\rho$ to construct an arbitrary teleportation
channel is given by
\be
d_{min}(\rho) = \frac{m}{m+1} \Delta(P_+, \rho).
\ee

{\em Proof}: $d_{min}(\rho) \leq m \Delta(P_+, \rho) /(m+1)$,
because a possible way to use $\rho$ as a teleportation channel is
by using a twirled version of an optimal state $P(\rho)$ and the
standard teleportation scheme, which produces a depolarizing
teleportation channel with $d=m D(P_+, P(\rho)) /(m+1)$ (recall
eq. (\ref{relation})). On the other hand $d_{min}(\rho)$ is at
least $m \Delta(P_+, \rho) /(m+1)$. Indeed, take an optimal
teleportation scheme employing the state $\rho$ and LOCC only. It
will produce some optimal teleportation channel $\Lambda$, that we
can turn into a depolarizing channel without increasing
$d(\Lambda_p^{dep})=d_{min}(\rho)$. Then we can send half of $P_+$
through the channel to obtain a noisy singlet $\rho_p$ that
satisfies eq. (\ref{relation}). The desired inequality follows
then from the fact that $D(P_+,\rho_p) \geq
\D(P_+,P_{opt}(\rho))$.

Therefore, using (\ref{bound-singlet}) we can announce the
following upper bound to the optimal teleportation distance
$d_{min}(\rho)$ achievable with state $\rho$ and LOCC

{\bf Proposition 6:}
\be
d_{min}(\rho) \geq \frac{2}{m+1}(m-1+2\N(\rho)).
\ee

The two results of this section can also be derived in terms of
fidelities (the so-called singlet and channel fidelities, see for instance \cite{Horox3-99}). The upper bounds one
obtains read,
 \bea
F_{opt} &\equiv& \max_P \bra{\Phi^+} P(\rho)\ket{\Phi^+}
          \leq \frac{1+2\N(\rho)}{m}; \\
f_{opt}(\rho) &\equiv& \max_{\Lambda_{\rho}}
     \int d\phi \bra{\phi} \Lambda(\proj{\phi})\ket{\phi}
     \leq\frac{2d(\N(\rho)+1)}{m+1}.
\eea

\section{Upper bound to distillation rates}

We now move to consider a second application of the previous measures, namely a bound on the asymptotic distillability of a mixed state $\rho$ in terms of $E_{\N}(\rho)$.

The distillation rate of a bipartite state $\rho$ is the best rate
at which we can extract near perfect singlet states from multiple
copies of the state by means of LOCC. The asymptotic (in the
number of copies) distillation rate is the so-called {\em
entanglement of distillation} $E_D(\rho)$ \cite{Be962}, one of the
fundamental measures of entanglement. In this section we will show
that the logarithmic negativity $E_{\N}$ is always at least as great as the
entanglement of distillation $E_D^{\epsilon}(\rho)$, where $\epsilon$ denotes the degree of imperfection allowed in the
distilled singlets.

Let $\Upsilon$ denote a maximally entangled state of two qubits,
and consider, for some number $n_{\alpha}$ of copies of $\rho$,
the best approximation to $m_{\alpha}$ copies of $\Upsilon$ that
can be obtained from $\rho^{\otimes n_{\alpha}}$ by means of LOCC.
Similarly to the previous section we define
 \be\label{Delta}
   \Delta(\Upsilon^{\otimes m_{\alpha}}, \rho^{\otimes n_{\alpha}})
     \equiv \inf_P\;\normt{\Upsilon^{\otimes m_{\alpha}}-P(\rho^{\otimes n_{\alpha}})}\;,
 \ee
where $P$ runs over all deterministic protocols built from LOCC.
We say that $c$ is an achievable distillation rate for $\rho$, if
for any sequences $n_\alpha,m_\alpha\to\infty$ of integers such
that $\limsup_\alpha(n_\alpha/m_\alpha)\leq c$ we have
 \be\label{Del2zero}
   \lim_\alpha\;\Delta(\Upsilon^{\otimes n_\alpha},
    \rho^{\otimes m_\alpha})=0\;.
 \ee
The distillable entanglement $E_D(\rho)$ corresponds then to the
supremum of all achievable distillation rates. Several variants of
this definition are available in the literature, which are
however, equivalent to the one given here. In particular, we may
replace `$\Delta\to0$' by a `fidelity$\to1$', and we may consider
selective protocols, in which operations produce variable numbers
of output systems on the same input, and the expected rate is
optimized. Of course, restricting the amount of classical
communication between Alice and Bob will in in general change the
rate.

The above definition requires that the errors go to zero, but in
many applications one can live with a small but finite error
level. Therefore we introduce $E_D^\epsilon(\rho)$, the distillable
entanglement at error level $\epsilon$, which is defined exactly
as above, but (\ref{Del2zero}) is replaced by
 \be\label{Del2eps}
   \limsup_\alpha\;\Delta(\Upsilon^{\otimes n_\alpha},
    \rho^{\otimes m_\alpha})\leq\epsilon\;.
 \ee
Of course, $E_D^0(\rho)=E_D(\rho)$, and $\epsilon\mapsto
E_D^\epsilon(\rho)$ is a non-decreasing function. The main result of
this section is the following bound.

{\bf Proposition 7:}
\be\label{upperB}
E_D^\epsilon(\rho)\leq E_\N(\rho)\;,
\ee
for all $0\leq\epsilon<1$.

{\em Proof:\/} The only property of LOCC operations used
in the proof is that for any such operation $P$, there is another,
$P'$ such that $P(\rho)\pt=P'(\rho\pt)$. We denote by $\Upsilon_d$
the maximally entangled state on a pair of $d$-dimensional spaces.
Then, as shown below, we have $\normt{\Upsilon_d\pt}=d$. In some
sense this is the worst case: for general hermitian operators we
have $\normt{A\pt}\leq d\normt A$.

Now suppose that $P$ is the transformation for which the infimum
(\ref{Delta}) for $\Delta(\Upsilon_d,\rho)$ is attained. Then
 \be\label{rhopt>}
   \normt{\rho\pt}\geq \normt{P'(\rho\pt)}=\normt{P(\rho)\pt}\;,
 \ee
where the first estimate holds, because $P'$, as a bona fide LOCC
operation, does not increase the trace norm (recall the
monotonicity of $\N(\rho)$). On the other hand,
 \bea\label{upsDelta}
   \normt{P(\rho)\pt}
     &\geq& \normt{\Upsilon_d\pt}- \normt{(\Upsilon_d-P(\rho))\pt}
         \nonumber\\
     &\geq& d- d\;\Delta(\Upsilon_d,\rho)\;.
 \eea
Taking the logarithm, we find
 \be\label{EN1shot}
  E_\N(\rho)\geq \log_2(d) + \log_2(1-\Delta(\Upsilon_d,\rho))\;.
 \ee
Now let $n_\alpha, m_\alpha$ be diverging integer sequences as in
the definition of achievable rate $c$. Then, using the additivity
of $E_\N$, and the last inequality with $d=2^{n_\alpha}$, we find
 \bea
  E_\N(\rho)&=& \frac1{m_\alpha}E_\N(\rho^{\otimes m_\alpha})
                 \nonumber\\
            &\geq& \frac1{m_\alpha}\Bigl(n_\alpha +\log_2
                \bigl(1-\Delta(\Upsilon^{\otimes n_\alpha},\rho^{\otimes
                      m_\alpha})\bigr)\Bigr)\;. \nonumber
 \eea
We now go to the limit superior with respect to $\alpha$,
observing that the error $\Delta$ is uniformly bounded away from
$1$, and $m_\alpha\to\infty$. Hence $E_\N(\rho)\geq c$ for every
achievable rate $c$, which concludes the proof.

\section{Explicit examples}

In this section we display explicit expressions for the negativity
for some particular classes of bipartite states, namely for arbitrary pure states, for mixed states with a high degree of symmetry, and finally also for Gaussian states of a light field.

\subsection{Pure states}

All entanglement measures based on asymptotic distillation and
dilution of pure-state entanglement, in particular the
entanglement of formation $E_F$ and the distillable entanglement $E_D$ 
\cite{Be962}, but also the relative entropy of entanglement
\cite{Ve98} agree on pure states, where they give the von Neumann
entropy of the restricted states. Negativity gives a larger value:
Let $\rho=\proj{\Phi}$ be a pure state, and write the wave vector
in its Schmidt decomposition $\Phi=\sum_\alpha c_\alpha
e'_\alpha\otimes e''_\alpha$, where $c_\alpha>0$ are the Schmidt
coefficients of $\Phi$, and the $e^{(i)}_\alpha$ are suitable
orthonormal basis. Then we get the following result.

{\bf Proposition 8:}
\be\label{NNNpure}
\N(\rho) = \frac12\Bigl(\bigl(\sum_\alpha c_\alpha\bigr)^2-1\Bigr)\;.
 \ee
This is precisely $\N_{ss}/2$, i.e. half of the robustness of
entanglement, as computed in \cite{rob}.

\noindent{\it Proof:\/}
Introducing the operators ``flip'' $\flip e'_\alpha\otimes
e''_\beta=e'_\beta\otimes e''_\alpha$, and $C'=\sum_\alpha
c_\alpha\ketbra{e'_\alpha}{e'_\alpha}$, and a similar $C''$ for
the second tensor factor, we find
\begin{eqnarray}
  \bigl(\ketbra\Phi\Phi\bigr)^{T_1}
  &=&\sum_{\alpha\beta}c_\alpha c_\beta\
    \ketbra{e'_\alpha\otimes e''_\beta}{e'_\beta\otimes e''_\alpha}
  \nonumber\\\label{ptpure}
  &=& \flip\;(C'\otimes C'')\;.
\end{eqnarray}
From the trace norm $\norm X_1=\tr\sqrt{X^{\dagger}X}$ we may omit unitary
factors like $\flip$, so the trace norm is equal to the trace of
the positive operator $(C'\otimes C'')$, namely $\bigl(\sum_\alpha
c_\alpha\bigr)^2$.

Since $E_\N$ is an upper bound on the distillation rate, and that
rate is known to be $E(\rho)$, the von Neumann entropy of the
restricted state, we know that $E_\N(\rho)\geq E(\rho)$. But of
course, we can get this more directly: using the concavity of the
logarithm, we get
 \be
  E(\rho)=2\sum_\alpha c_\alpha^2\log_2(\frac1{c_\alpha})
      \leq2\log_2\Bigl(\sum_\alpha c_\alpha\Bigr)
      =E_\N(\rho)\;.
 \ee
This derivation also allows the characterization of the cases of
equality: Since the logarithm is strictly concave, we get equality
if and only if all non-zero $c_\alpha$ are equal. Hence equality
for pure states holds exactly for maximally entangled states
(which may have been expanded by zeros to live on a larger Hilbert
space).

\subsection{States with symmetry}
All entanglement measures can be computed more easily for states that
are invariant under some large group of local unitary
transformations \cite{VW,Rains}. The negativity is no exception.
The main gain from local symmetries is that the partial transpose
lies in a low dimensional algebra, and is hence easily
diagonalized. For this background we refer to \cite{VW}. But often
a direct computation is just as easy.

Consider, for example, the states $\rho$ on $\C^d\otimes\C^d$,
which commute with all unitaries of the form $U\otimes U$, where
$U$ is real orthogonal. These can be written as
 \be\label{OOstate}
  \rho=ad\proj{\Phi^+}+ b \flip +c \idty\;,
 \ee
where $\ket{\Phi^+}=(\sum_{\alpha=1}^d \ket{\alpha \otimes \alpha})/\sqrt{d}$ is again the
standard maximally entangled vector, and $a,b,c$ are suitable real
coefficients. This family includes both the so-called Werner
states \cite{werner} with $a=0$ and, with $b=0$, the so-called
isotropic states \cite{iso} (or noisy singlets, compare
Eq.~(\ref{noising}) above). The three operators in this expansion
commute, so all operators of the form (\ref{OOstate}) can be
diagonalized simultaneously, with spectral projections
 \bea
   p_0&=&\proj{\Phi^+}  \nonumber\\
   p_1&=&(\idty-\flip)/2  \nonumber\\
   p_2&=&(\idty+\flip)/2-\proj{\Phi^+}\;.  \nonumber
 \eea
We parameterize the states of the form (\ref{OOstate}) by the two
expectation values $f=d \tr(\rho \proj{\Phi^+})$ and
$g=\tr(\rho\flip)$, the third parameter for determining $a,b,c$
being given by the normalization. Then the states correspond to
the triangle $0\leq f\leq d$, $-1\leq g\leq 1$, $f\leq d(1+g)/2$.

Since partial transposition simply swaps the operators $\flip$ and
$d\proj{\Phi^+}$, leaving $\idty$ unchanged, we can apply the same
method to compute the trace norm of the partial transpose, and
hence $\N(\rho)$. Explicitly, we get
 \be\label{Ns4oo}
   \N(\rho)
        =\frac14\abs{1-f}+\frac14\abs{1+f-2g/d}+\frac12\abs{g/d}-\frac12
 \ee

It turns out \cite{VW} that in this class of states the
Peres-Horodecki separability criterion holds (in spite of
arbitrary dimension $d$), i.e., the set of ppt-states is the same
as the set of separable states, and in the parameterization chosen
equal to the square $f,g\in[0,1]$. Hence
$\N_{ss}(\rho)=\N_{ppts}(\rho)$. Evaluating a simple variational
expression we get
 \be\label{Nssoo}
   \N_{ss}(\rho)=\frac12\max\Bigl\lbrace \abs{2f-1}-1,
                    \abs{2g-1}-1,0\Bigr\rbrace\;.
 \ee

\subsection{Gaussian states}
Gaussian states frequently occur in applications in quantum
optics, where they describe the light field. Alice's and Bob's
systems are then described by a certain number of canonical
degrees of freedom such as field quadratures of suitable modes.
However, the same formalism applies when the canonical operators
are positions and momenta of a certain number of harmonic
oscillators. Gaussian states are then defined as those (possibly
mixed) states with Gaussian Wigner function.

For simplicity we denote the full collection of canonical
operators by $R_\alpha,\alpha=1,\ldots,2n$, where Alice holds
$n_A$ oscillators and Bob holds $n_B$, and $n=n_A+n_B$. These are
either position or momentum operators, whose commutation relations
are of the form
 \be\label{ccr}
  [R_\alpha,R_\beta]=i\sigma_{\alpha\beta}\idty\;,
 \ee
with an antisymmetric scalar matrix $\sigma$, called the
symplectic matrix, which has the block matrix decomposition
 \be
\Delta =\left(
\begin{array}{c|c}
   \sigma_A & 0\\ \noalign{\hrule} 0&\sigma_B
\end{array}\right)\ee
with respect to a decomposition of the set of indices into Alice's
and Bob's. This form expresses the fact that all variables of
Alice commute with all of Bob's.

A Gaussian state is determined by its first two moments
$m_\alpha=\tr(\rho R_\alpha)$ and
 \be\label{moment2}
  \gamma_{\alpha\beta}
    =\tr(\rho R_\alpha R_\beta)-\frac i2\sigma_{\alpha\beta}\;,
  \ee
where the subtraction is chosen as the antisymmetric part of
$\tr(\rho R_\alpha R_\beta)$, which is fixed by the commutation
relations, independently of the state. $\gamma$ is then a real
symmetric matrix. Since the mean $m_\alpha$ can be made zero by a
local unitary transformation (a translation in phase space) it is
irrelevant for entanglement, and we choose it to be zero. The
second moment $\gamma$ is then the same as the {\it covariance
matrix} of the state. The uncertainty relation, the universal
lower bound on variances, is then expressed as the positive
definiteness of $\gamma_{\alpha\beta}+(i/2)\sigma_{\alpha\beta}$.
It will be crucial for later that for a classical Gaussian it is
only necessary that $\gamma$ itself is positive definite. Hence we
can have non-positive operators, whose Wigner function is an
ordinary, if somewhat sharply peaked Gaussian.

In order to compute the trace norm of such an operator or, more
generally, to compute the spectrum or other characteristics not
depending on the Alice|Bob partition of the system, we can bring
$\gamma$ into a standard form by a process known as symplectic
diagonalization or {\it normal mode decomposition}. This means
choosing a suitable canonical linear transformation (i.e., a
transformation leaving the symplectic form $\sigma$ invariant),
which can be implemented on the Hilbert space level by unitary
operators (known as the metaplectic representation). Assuming
$\sigma$ to be in standard form, i.e., block diagonal with $n$
$2\times2$ blocks of the form
 $\Bigl(\begin{array}{cc} 0&1\\1&0\end{array}\Bigr)$
this results in a diagonal $\gamma$, with equal
eigenvalues for each block, i.e. $\gamma={\rm
diag}(c_1,c_1,c_2,c_2,\ldots,c_n,c_n)$. We call
$(c_1,c_2,\ldots,c_n)$ the {\it symplectic spectrum} of $\gamma$.
The fast way to compute it is via the eigenvalues of the matrix
$\sigma^{-1}\gamma$, which are $\pm ic_1,\ldots,\pm ic_n$. At the
Hilbert space level the normal mode decomposition transforms the
state into a tensor product of independent harmonic oscillators,
each of which is in a thermal oscillator state, the temperature
being a function of $c_\alpha$. The smallest value allowed by
uncertainty is $c_\alpha=1/2$, which gives the oscillator ground
state.

In this context {\it transposition} is best identified with time
reversal. Indeed, in the general scheme we can choose a basis in
which the transpose is computed, but all these choices are
equivalent via a local unitary transformation. In this case we
choose the position representation, in which transposition is the
same as reversing all momenta, keeping all positions, and to lift
this to products by observing that transposition reverses operator
products. Partial transposition $T_A$ is completely analogous.
Only in this case just Alice's momenta are reversed and Bob's are
left unchanged. On the level of Wigner functions and covariance
matrices of Gaussians, we just have to apply the corresponding
linear transformations on phase space. That is, $\gamma^{T_A}$,
the covariance matrix of the partial transpose of a Gaussian state
with covariance matrix $\gamma$ is constructed by multiplying by
$-1$ all matrix elements, which connect one of Alice's momenta to
either a position, or a degree of freedom belonging to Bob, and
leaving all other matrix elements unchanged.

The point is, of course, that while this transformation preserves
the positive definiteness of $\gamma$, and hence we get another
Gaussian Wigner function, it does {\it not} respect the
uncertainty relation, so the partially transposed operator may
fail to be positive. However, the whole formalism of normal mode
decomposition for $\gamma$ works exactly as before: we get a
representation of the partial transpose as a tensor product of
(not necessarily positive) trace class operators. The trace norm
of this operator is just the product of the trace norms, so we
have completely reduced the computation of the trace norm to the
single mode case. To summarize the results so far:

\par{\it Let $\rho$ be a Gaussian density operator with
covariance matrix $\gamma$, and let $(\widetilde
c_1,\ldots,\widetilde c_n)$ be the symplectic spectrum of
$\gamma^{T_A}$. Then
 \be\label{Egauss}
 E_\N(\rho)=\sum_{\alpha=1}^n F(\widetilde c_\alpha)\;,
 \ee
where $F(c)=\log_2\norm{\rho_c}_1$, and  $\rho_c$ is the operator
whose Wigner function is a Gaussian with covariance ${\rm
diag}(c,c)$.}

Of course, the function $F$ vanishes for $c\geq(1/2)$. It is
easily determined by looking at Gaussian states for oscillators as
the temperature states of the oscillator. With $z=e^{-\beta}$, and
$\ket n$ the $n$th eigenstate of the oscillator, a general
Gaussian is of the form
 \be\label{gaussZ}
   \rho=(1-z)\sum_{n=0}z^n\;\ketbra nn\;
 \ee
where $z\ge0$ corresponds to density operators, and $-1<z<0$ to
Gaussians whose Wigner functions have sub-Heisenberg variance.
Then we get
 \bea\label{gaussZ2}
   \norm\rho_1&=&(1-z)(1-\abs z)^{-1}\\
   c&=&\tr(\rho P^2)=\tr(\rho\frac12(P^2+ Q^2))\nonumber\\
    &=&(1-z)\sum_nz^n\;(n+\frac12)\nonumber\\
    &=& (1-z)^{-1}-\frac12\;\label{gaussZ3}
 \eea
Solving Eq. (\ref{gaussZ3}) for $z$ and substituting into
(\ref{gaussZ2}) we find
 \be\label{Fgauss}
   F(c)=\left\lbrace\begin{array}{cl}
              0           &\hbox{for } 2c\geq1\\
              -\log_2(2c) &\hbox{for } 2c<1
        \end{array}\right.
 \ee
Together with (\ref{Egauss}) and the process of normal mode
decomposition this is an efficient procedure for determining
$E_\N(\rho)$.

In the simplest case of one oscillator each for Alice and Bob we
may go even further, by expressing the symplectic spectrum of the
partial transpose directly in terms of the covariance matrix.
Suppose that
 \be\label{1osci}
   \gamma=\left(
              \begin{array}{c|c}
                A & C\\ \noalign{\hrule} C^T&B
              \end{array}\right) \;,
 \ee
with $2\times 2$ matrices $A,B,C$. Then, as shown in \cite{SimonR}, the numbers $\det A,\;\det B,\; \det C\;$, and
$\det\gamma$ are a complete set of invariants for $\gamma$ with
respect to local symplectic transformations. Moreover, when
passing from $\gamma$ to $\gamma\pt$ only $\det C$ changes sign,
and the others remain unchanged. Pure states are characterized by
the conditions $\det\gamma=1/16$, and $\det A+\det B+2\det C=1/2$,
and can be brought into the normal form
 \be\label{1oscipure}
   \gamma=\left(
              \begin{array}{cc|cc}
                a&0&c&0\\0&a&0&-c\\ \noalign{\hrule}
                c&0&a&0\\0&-c&0&a
              \end{array}\right) \;,
 \ee
where $a^2=c^2+1/4$.

Coming back to the general case of (\ref{1osci}), the
characteristic equation of $\sigma^{-1}\gamma\pt$, whose solutions
are the $\pm\widetilde c_\alpha$, takes the form
 \be \xi^4+(\det A+\det B-2\det C)\xi^2+\det\gamma=0
 \ee
Together with Eq.~(\ref{Egauss}) this amounts to an explicit
formula. For the particular case of a pure state we find
 \be
  E_\N(\rho)=-2\log_2(\sqrt{a-1/2}-\sqrt{a+1/2})\;,
 \ee
which is readily seen to agree with Eq.~(\ref{NNNpure}).

\section{Multipartite systems}

 As argued in the introduction, a computable measure of entanglement 
for bipartite mixed
states is also very convenient for the quantification of
multipartite entanglement. In this section we describe a whole set
of computable parameters related to the negativity that can be
associated to a multipartite state to make quantitative statements
about its entanglement.

\subsection{Multipartite negativities}

Consider a quantum system consisting of, say, three parts,
associated to Alice, Bob and Charlie. Let $\rho_{ABC}$ be the
(either pure or mixed) state of the system. A possible way to
classify the entanglement properties of such a state is by looking
at the different bipartite splittings \cite{spli} of the system.

First, we can join two of the three parts, say those of Alice and
Bob, and compute the sum of negative eigenvalues of
$\rho_{ABC}^{T_C}$, $\N_{(AB)-C}(\rho_{ABC})$. This is
automatically an entanglement monotone \cite{multimono}, which quantifies the
strength of quantum correlations between Charlie and the other two
parties. Similarly, the negativities $\N_{(AC)-B}(\rho_{ABC})$ and
$\N_{(BC)-A}(\rho_{ABC})$ are two other monotonic functions under
LOCC with analogous meaning.

We can also consider the entanglement properties of two-party
reduced density matrices. Suppose, for instance, that Charlie
decides not to cooperate with the two other parties in the
manipulation of the tripartite system according to LOCC. Alice and
Bob's effective density matrix, $\sigma_{AB} \equiv
$Tr$_C[\rho_{ABC}]$, may still retain some of the original
entanglement. The negativity of $\sigma_{AB}$,
$\N_{A-B;\not{C}}(\rho_{ABC})$, can be used to quantify this
residual entanglement. Analogous quantities can be used to
quantify the entanglement of $\sigma_{AC}$ and  $\sigma_{BC}$.

Thus, altogether we have obtained $6$ computable functions to
quantify the entanglement of any state of a tripartite system. In
a four-partite setting the number of possible splittings is much
greater (see \cite{spli} for a more detailed description), and
thus, we obtain up to $26$ inequivalent measures, namely: ($i$) $\N_{A-BCD}(\rho_{ABCD})$ and the corresponding permutations, i.e. $4$ inequivalent measures; ($ii$) $\N_{AB-CD}(\rho_{ABCD})$ and permutations ($4$ measures); ($iii$) $\N_{A-BC;\not{D}}(\rho_{ABCD})$ and permutations ($12$ measures); ($iv$) $\N_{A-B;\not{C}\not{D}}(\rho_{ABCD})$ and permutations ($6$ measures).

\subsection{Hierarchy}

Notice that although all these measures are independent functions of the multipartite state, there is a strength hierarchy between them when corresponding to related bipartite splittings with different number of parties. In the four-party case we have that, for instance,
\be
\N_{A-BCD} \geq \N_{A-BC;\not{D}} \geq \N_{A-B;\not{C}\not{D}},
\ee
which follows from the fact that to trace out a part of a local
system is an operation of the set LOCC, under which the negativity can only decrease. Of course, the same inequalities hold for the corresponding logarithmic negativities, and thus also for the several bounds on distillability ---of different kinds of multipartite entanglement--- implied by the later. 

It should be noted, however, that in this way one can quantify
only some aspects of multi-particle entanglement: there are
tripartite states which are separable with respect to every
splitting of the system, but are nevertheless not a convex
combination of triple tensor products of density operators
\cite{spli,tripartite}. States which have positive partial transpose
with respect to every subsystem satisfy a large class of Bell
inequalities \cite{Bellmulti}.

\section{Discussion and conclusions}

In this paper we have presented a computable measure of
entanglement for bipartite mixed states, the negativity $\N(\rho)$,
which we have proved not to increase under LOCC. Although it lacks
a direct physical interpretation, we have shown that it bounds two
relevant quantities characterizing the entanglement of mixed
states: the channel capacity and the distillable entanglement
$E_D^{\epsilon}$.

Ideally, quantum correlations would be best quantified by measures
with a given physical meaning. Which measure to be used, would
depend on which question we want to answer. For instance, if we
want to know how much pure-state entanglement the parties can
extract from (infinitely) many copies of the state $\rho$, then
the proper measure to be used is the entanglement of distillation
$E_D(\rho)$. 

In practice, however, the value of these measures is
not known. Recent studies of entangled systems, such as those of
{\em entangled chains}, {\em entanglement molecules}, {\em
entangled rings}, {\em entangled Heisenberg models} and {\em cluster states}
\cite{exemples}, which are $N$-qubit systems in some global
entangled state, are focused on the two-qubit quantum
correlations associated to the global state, as measured by the
entanglement of formation (or the related concurrence)
\cite{Wo98}. This choice of measure of entanglement is somehow
arbitrary --- where not simply forced by the lack of an
alternative measure that can also be computed for two-qubit mixed
states---, because it does not reflect anyway the entanglement
cost of formation of the $N$-qubit state. 
We envisage that in these and similar
contexts it will pay off to use a computable entanglement
measure like the negativity, whose evaluation is not restricted to
two-qubit mixed states. The negativity will allow, for instance, to
generalize the previous investigations to analogous constructions with 
$l$-level systems ($l>2$) instead of qubits, also to analyze 
quantitatively the entanglement between subsets of these $l$-level 
systems.

 Finally, in a similar way as the negativity has recently played a role in proving the irreversibility of asymptotic local manipulation of bipartite mixed-state entanglement \cite{Vidal}, we hope that this computable measure will also be a useful tool to answer other fundamental questions of entanglement theory.

\section*{Acknowledgments}

G.V. acknowledges a Marie Curie Fellowship HPMF-CT-1999-00200
(European Community). This work was also supported by the project
EQUIP (contract IST-1999-11053), the European Science Foundation,
and the DFG (Bonn).


\begin{references}
\bibitem{dagger}The notation for adjoints and complex conjugates
of matrices is unfortunately not uniform in the literature. The
notations used here are those usually employed for the first
author, i.e., $U^\dagger$ for the adjoint of $U$, and $U^*$ for
the matrix element-wise complex conjugate. The second author would
normally follow the usage of another branch of the physics
literature, as well as a majority of mathematicians, using $U^*$
for the adjoint, and $\overline{U}$ for the complex conjugate. It
seems the quantum information community will have to live with
this ambiguity.
\bibitem{multimono} Notice that since the quantity $\N_{(AB)-C}$ is a monotonic function of the state a bipartite system $(AB)-C$ under LOCC, it is also a monotone under LOCC in the tripartite setting $A-B-C$, where local transformations are additionally restricted by the fact that $A$ and $B$ are different parties.
\end{references}
\end{document}